\documentclass[10pt,twoside,twocolumn]{article}
\usepackage{balance} 

\usepackage[utf8]{inputenc}
\usepackage[T1]{fontenc}

\usepackage[margin=1in]{geometry}

\usepackage{amsmath,amssymb,amsfonts}
\usepackage{mathtools}

\usepackage{graphicx}
\usepackage{subcaption} 

\usepackage{cite}
\usepackage[hidelinks]{hyperref}

\usepackage{authblk}

\newcommand{\ket}[1]{\lvert #1\rangle}

\def\BibTeX{{\rm B\kern-.05em{\sc i\kern-.025em b}\kern-.08em
    T\kern-.1667em\lower.7ex\hbox{E}\kern-.125emX}}

\newcommand{\keywords}[1]{%
  \par\noindent\textbf{Index Terms—}#1\par
}

\newcommand{\PARstart}[2]{#1#2}

\title{Resource-Efficient Emulation of Majorana Zero Mode Braiding on a Superconducting Trijunction}

\author[1]{Rahul Singh}
\author[2]{Weixin Lu}
\author[3]{Kaelyn Ferris}
\author[2,$\dagger$]{Javad Shabani}

\affil[1]{Department of Electrical and Computer Engineering, University of California, Santa Barbara, CA 93106, USA}
\affil[2]{Center for Quantum Information Physics, Department of Physics, New York University, New York, NY 10003, USA}
\affil[3]{IBM Quantum, IBM T.~J.~Watson Research Center, Yorktown Heights, NY 10598, USA}
\affil[$\dagger$]{Corresponding author: Javad Shabani (\texttt{jshabani@nyu.edu})}

\date{} 

\begin{document}
\twocolumn[
\maketitle

\begin{abstract}
Topological superconductivity could host quasiparticles that are key candidates for fault-tolerant quantum computation due to their immunity to noise as they obey non-Abelian exchange statistics. For example, in the case of Majorana Zero Modes (MZM), braiding enables two topologically protected quantum gates. While their direct manipulation in solid-state systems remains experimentally challenging, digital emulation of MZM behavior has provided insight as well as a deeper understanding of controlling these topological quantum systems. This emulation is typically accomplished by mapping the topological and trivial phases of a Majorana system to ferromagnetic and paramagnetic Hamiltonians of a spin-glass model. This approach usually relies on adiabatic evolution of superconducting Hamiltonians, which require circuits with very large depths. In this work, we present a resource-efficient method to emulate MZM braiding in a trijunction geometry using a quantum processor. We introduce direct braiding operators which simulate the evolution more efficiently, reducing the quantum gate overhead. We then further generalize this method to emulate braiding operations in extended trijunction architectures based on Kitaev chains.
\end{abstract}

\vspace{0.75em}
]

\section{Introduction}
\label{sec:introduction}
\PARstart{I}{n} topological superconductors, Majorana Zero Modes (MZMs) obey non-Abelian statistics~\cite{Kitaev_2001, Alicea2011, Beenakker2020, Clarke2011}, enabling quantum information to be stored non-locally and thereby protected from local perturbations and disorder~\cite{Will_towards_topo_supercond_2022, Bassel_planar_JJ_2024, Lutchyn2010, Oreg2010, Klinovaja2012, Bassel_mag_2023, Ramis_Movassagh_2020, Aasen2016}. This inherent robustness makes MZMs particularly attractive for fault-tolerant quantum computation~\cite{Nayak2008, Aasen2016, DasSarma2015, Flensberg_2021, Bedow_2024}.

Several studies have investigated the simulation of Majorana-like behavior across diverse topologies~\cite{Bassel_ising_exchange_2021,Backens_2017, Bomantara2017} and its bosonic analog~\cite{Busnaina_2024}, with some emulations based on quantum dots~\cite{Xu2023} and quantum computers and simulators~\cite{Harle2023, Stenger2022}. Of particular interest is the braiding of MZMs in a trijunction, which serves as the logical operation for a topological superconducting qubit device, making the emulation of Majorana-like behavior critical for large systems~\cite{Marek_2021, Harle2023, Mi_2022}. Simulating these systems using superconducting quantum computers has also been explored~\cite{Stenger_2021, You_2014, Sung_2023}, although these approaches typically rely on adiabatic evolution, which often results in large quantum circuits—posing challenges for near-term intermediate-scale quantum (NISQ) devices~\cite{Stenger_2021}. Due to current hardware limitations, such simulations are often noisy and restricted in scale.

\begin{figure}[t!]
  \centering
  \includegraphics[width=0.99\linewidth]{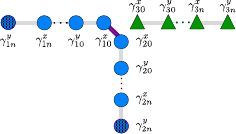}
  \caption{Pictorial representation of a topological trijunction. The three arms of the device are shown as gray bars. Majorana modes in the trivial phase are depicted as green triangles, while Majorana modes in the topological phase are shown as blue circles. The shaded circles indicate the unpaired, non-local Majorana modes. The purple line represents the coupling between two topological-phase arms, with the remaining arm in the trivial phase.}
  \label{fig:TriJunction}
\end{figure}

To emulate the braiding of Majorana zero modes (MZMs), we consider a superconducting topological trijunction, shown in Fig.~\ref{fig:TriJunction}. Each wire consists of $n$ lattice sites and, depending on whether it is tuned to the topological phase, can host an end-localized MZM. In the topological regime, pairs of end Majoranas define a non-local fermionic mode. Such trijunction architectures have been studied extensively, both for hardware implementations and for digital/analog simulation of Majorana dynamics~\cite{Stenger_2019, Stenger_2021}. We model each arm as a Kitaev chain with nearest-neighbor couplings and include tunable couplings at the junction to control interactions between them.

The Kitaev chain, a prototypical system that contains MZMs, when in its topological phase, hosts two unpaired Majorana fermions at the chain’s ends; however, in the trivial phase, such modes are absent. These Majorana fermions act as their own antiparticles (i.e., $\gamma = \gamma^\dagger$). The general Hamiltonian for a Kitaev chain is given by:
\begin{equation}
    \begin{split}
        H &= \frac{i}{2}  \sum_j \bigl( -\mu\gamma_{j}^x\gamma_{j}^y + (t + |\Delta|) \gamma_{j}^y\gamma_{j+1}^x \\
        &\quad + (-t+|\Delta|) \gamma_{j}^x\gamma_{j+1}^y \bigr),
    \end{split}
    \label{eq:KitaevsChain}
\end{equation}
where $\mu$ is the on-site energy, $\Delta$ is the induced superconducting gap, and $t$ is the hopping amplitude. In this work, we consider the regime $\Delta = t$. In this regime for the limiting case, the Kitaev chain has two distinct phases:
\begin{itemize}
    \item \textbf{Trivial phase:} $t = 0$, $\mu > 0$
    \item \textbf{Topological phase:} $t > 0$, $\mu = 0$.
\end{itemize}

In this work, we propose an alternative method to simulate braiding in a quantum computing environment which significantly reduces the circuit depth required to emulate Majorana-like behavior. Furthermore, we extend this approach to generalize the method for trijunctions of arbitrary length. In particular, our main contributions are as follows:
\begin{itemize}
    \item We construct an explicit braiding operator protocol for a topological trijunction and show that it implements the same trijunction-to-trijunction transformations as an adiabatic braiding path, while acting directly in the Majorana operator algebra.
    \item We develop qubit mappings tailored to this geometry, using a coupler-based Jordan-Wigner transformation, and derive the corresponding qubit Hamiltonians and Pauli-string decompositions suitable for current gate-based quantum hardware.
    \item We provide a detailed resource analysis comparing adiabatic and braiding-operator implementations, demonstrating that the latter achieves substantially lower two-qubit gate counts and circuit depths, with favorable scaling in the length of the trijunction arms.
    \item We validate the protocol on finite trijunctions of increasing size and identify the regimes in which the braiding-operator approach is most advantageous for near-term quantum emulation experiments.
\end{itemize}

The remainder of this paper is organized as follows: Section~\ref{Sec:TriJunction} introduces the topological superconducting trijunction, the associated braiding protocol, the mappings from Majorana operators to Pauli operators, and the emulation schemes we consider. Section~\ref{Sec:Result} presents implementation details, circuit-level resource estimates, and numerical results that validate the proposed braiding-operator-based emulation.

\section{Braiding in a Topological Trijunction}
\label{Sec:TriJunction}

We consider a superconducting trijunction whose three arms are labeled by $a,b,c \in \{1,2,3\}$, with sites along each arm indexed by $j = 0,\dots,n-1$. At each site $j$ of arm $k \in \{a,b,c\}$ we introduce two Majorana operators, $\gamma_{kj}^x$ and $\gamma_{kj}^y$, which  obey
\begin{equation}
    \bigl(\gamma_{kj}^\mu\bigr)^\dagger = \gamma_{kj}^\mu,\qquad
    \{\gamma_{kj}^\mu,\gamma_{k'j'}^{\nu}\}
    = 2\,\delta_{kk'}\delta_{jj'}\delta_{\mu\nu},
\end{equation}
for $\mu,\nu \in \{x,y\}$. The braiding protocol starts with two unpaired Majoranas located at the ends of arm 1 and arm 2, respectively, with both arms in the topological phase, while the third arm remains in the trivial phase. During the braiding process, the arms cyclically transition between topological and trivial phases, causing the two MZMs to move through different sites of the trijunction.

\subsection{Braiding Protocol in a Trijunction}
\label{Sec:TriJunctionProtocol}

\begin{figure}[t!]
  \centering
  \includegraphics[width=0.99\linewidth]{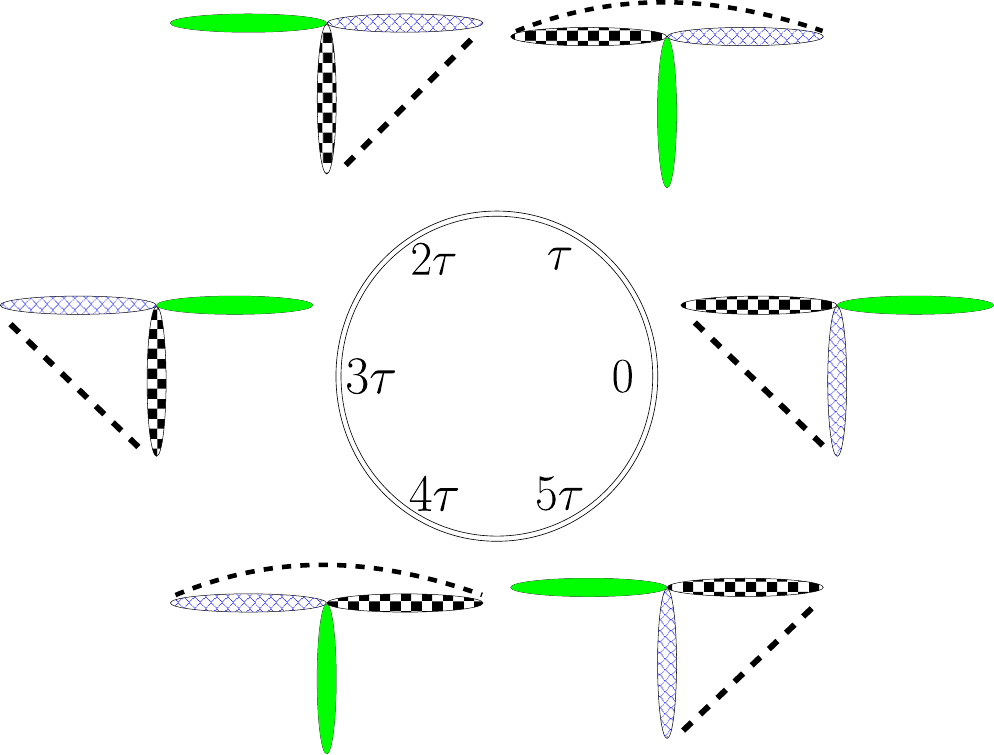}
  \caption{Time evolution of the braiding protocol starting at $t=0$, with time increasing counterclockwise around the trijunction. Arms connected by a dashed line are in the topological phase, while the unshaded arm is in the trivial phase. The trijunction sequentially passes through six distinct configurations at discrete time steps, each of duration $\tau$.}
  \label{fig:BraidingProtocol}
\end{figure}

The dynamics of the trijunction are governed by three Hamiltonians corresponding to the three possible choices of topological arms. The Hamiltonian for the initial configuration, $H_{12}$, combines the topological and trivial Kitaev chain Hamiltonians~\cite{Kitaev_2001,Sau2011}, with $a=1$ and $b=2$. As the system is evolved into other configurations, the Hamiltonian is similarly obtained by reassigning $a$ and $b$ to the topological arms and $c$ to the trivial arm:
\begin{equation}
    \begin{split}
        H_{ab} &=  i \Delta \sum_{k \in \{a, b\}} \sum_{j=0}^{n-2} \gamma_{kj}^y\gamma_{k,j+1}^x
        + i \alpha \sum_{l=0}^{n-1} \gamma_{cl}^x\gamma_{cl}^y\\
        &\quad + i \epsilon_{abc} t_{ab} \gamma_{a0}^x\gamma_{b0}^x,
    \end{split}
    \label{eq:MZMEquation}
\end{equation}
where $\Delta$ is the induced superconducting gap, $\alpha = -\mu/2$ encodes the on-site energy in the trivial arm, $t_{ab}$ denotes the coupling at the junction between arms $a$ and $b$, and $\epsilon_{abc}$ is the fully antisymmetric Levi--Civita symbol with $\epsilon_{123} = +1$. The braiding protocol, illustrated in Fig.~\ref{fig:BraidingProtocol}, consists of six steps in which the pattern of topological and trivial arms cycles around the trijunction. Among these six steps, three distinct phase configurations occur, corresponding to which pair of arms is in the topological phase. As the system size $n$ increases, the unpaired Majorana must traverse an arm to the junction in order to be transferred to a different arm.

\subsection{Mapping to Qubit Operators}
\label{Sec:Mapping}

In the six-step braiding protocol used here, $\tau$ denotes the evolution time per step, so that the total braiding time is $6\tau$. To emulate this process on a gate-based quantum processor, the trijunction must first be mapped onto the hardware qubits. Once such a mapping is chosen, all Majorana operators appearing in the Hamiltonians are converted to Pauli operators via an appropriate transformation.

In this work we employ a modified, coupler-based Jordan--Wigner (JW) transformation, which introduces an auxiliary qubit $C$ acting as a coupler. This auxiliary qubit keeps the resulting operators local and facilitates interactions at the junction between arms~\cite{Backens_2017,Stenger_2021}. The JW transformation is chosen so as to originate from this coupler qubit for each arm, effectively partitioning the trijunction into three independent Kitaev chains with the coupler acting as a shared site. This modified JW transform for arm $a$ and coupler $C$ is
\begin{equation}
    \begin{split}
        \gamma^x_{aj} &= \sigma_C^{(a)} \,\sigma_{aj}^x
            \prod_{i=0}^{j-1} \sigma_{ai}^z, \\
        \gamma^y_{aj} &= \sigma_C^{(a)} \,\sigma_{aj}^y
            \prod_{i=0}^{j-1} \sigma_{ai}^z,
    \end{split}
    \label{eq:CouplerTransform}
\end{equation}
where $\sigma_{aj}^\mu$ denotes the Pauli operator along axis $\mu \in \{x,y,z\}$ acting on site $j$ of arm $a$, and $\sigma_C^{(a)}$ denotes the Pauli operator acting on the coupler qubit along the axis associated with arm $a$. Since the three arms are labeled by $a \in \{1,2,3\}$, we identify $\sigma_C^{(1)} \equiv \sigma_C^x$, $\sigma_C^{(2)} \equiv \sigma_C^y$, and $\sigma_C^{(3)} \equiv \sigma_C^z$.

Using~\eqref{eq:CouplerTransform}, the Hamiltonians in Eq.~\eqref{eq:MZMEquation} map to
\begin{equation}
    \begin{split}
        H_{ab} &=
        -\Delta \sum_{k \in \{a,b\}} \sum_{j=0}^{n-2}
            \sigma_{kj}^x \sigma_{k,j+1}^x \\
        &\quad - \alpha \sum_{l=0}^{n-1} \sigma_{cl}^z \\
        &\quad - \epsilon_{abc}\, t_{ab}\,
            \sigma_C^{(c)} \,\sigma_{a0}^x \sigma_{b0}^x,
    \end{split}
    \label{eq:CouplerHamiltonian}
\end{equation}
where $\sigma_C^{(c)}$ denotes the Pauli operator on the coupler $C$ along the axis associated with the remaining arm $c$. In this mapping, the three arms $a$, $b$, and $c$ and the coupler $C$ can be arranged with the coupler located at the junction of the three-way topology. For comparison, we describe mapping the system using the standard JW transformation in Appendix~\ref{App:JW-appendix}.

\subsection{Adiabatic Emulation}
\label{Sec:Adiabatic}

\begin{figure*}[!t]
  \centering

  \begin{subfigure}{0.99\linewidth}
    \centering
    \includegraphics[width=0.99\linewidth]{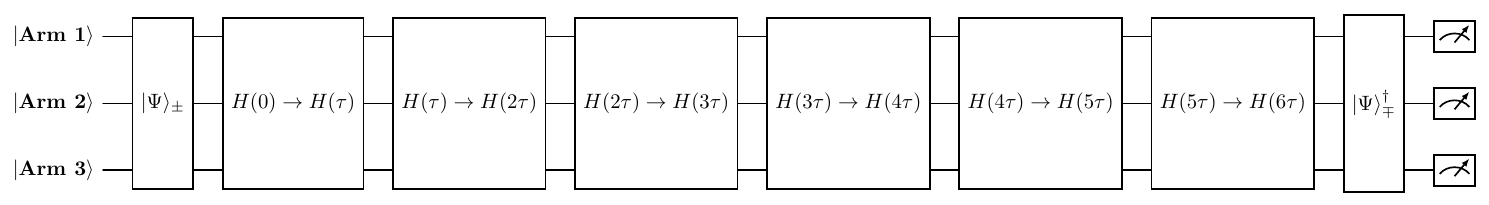}
    \caption{\scriptsize Adiabatic evolution: Each block labeled $H(I)\rightarrow H(F)$ denotes adiabatic evolution under a time-dependent Hamiltonian interpolating between an initial Hamiltonian $H(I)$ and a final Hamiltonian $H(F)$ over a time interval $\tau$. The full braid is realized by concatenating six such intervals of duration $\tau$.}
    \label{fig:braid-adiabatic}
  \end{subfigure}

  \begin{subfigure}{0.99\linewidth}
    \centering
    \includegraphics[width=0.99\linewidth]{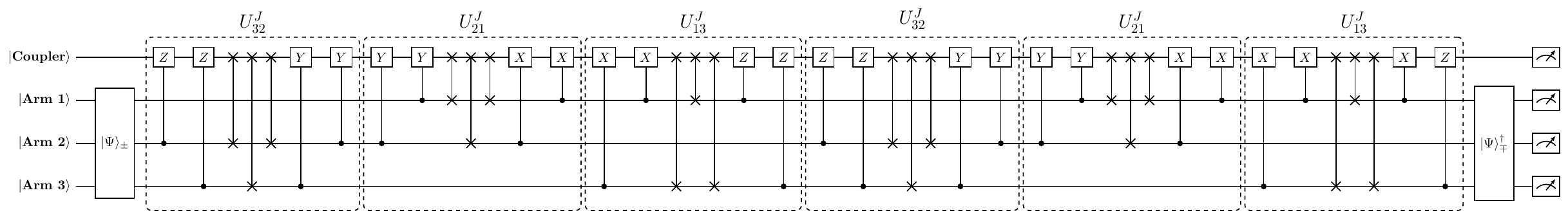}
    \caption{\scriptsize Braiding operators: Each block labeled $U^{J}_{ab}$ denotes a junction braiding unitary for the single-site trijunction. The sequence of six such operators implements the same logical braid as in Fig.~\ref{fig:braid-adiabatic}, using the operator-based construction introduced in Sec.~\ref{Sec:Braiding}.}
    \label{fig:braid-operator}
  \end{subfigure}

  \caption{Quantum circuits implementing the trijunction braiding protocol for a single-site trijunction. Panel~\ref{fig:braid-adiabatic} shows an adiabatic realization of the six-step protocol, while panel~\ref{fig:braid-operator} shows the corresponding braiding operator based construction. The blocks labeled $|\Psi\rangle_\pm$ and $|\Psi\rangle_\pm^\dagger$ denote the initial and final state preparation respectively for each implementation.}
  \label{fig:braid-circuits}
\end{figure*}

Braiding is widely understood as the physical realization of a fundamental quantum logic operation within a topological superconducting gap~\cite{Marek_2021,Harle2023}. The system is initialized in the state $|\Psi(0)\rangle_\pm$, which is a superposition of the two lowest-energy ground states, $|g_1\rangle$ and $|g_2\rangle$:
\begin{equation}
    |\Psi(0)\rangle_\pm = \frac{|g_1\rangle \pm |g_2\rangle}{\sqrt{2}}.
\end{equation}
After the completion of the six-step braiding protocol, at time $6\tau$, the system evolves into the complementary superposition of the same states:
\begin{equation}
    U_{BP} |\Psi(0)\rangle_\pm = |\Psi(6\tau)\rangle_\pm = |\Psi(0)\rangle_\mp,
\end{equation}
where $U_{BP}$ is the operator implementing the braiding process, and $|\Psi(t)\rangle$ denotes the state of the system at time $t$. The precise form of the lowest-energy eigenstates depends on the system size and the chosen mapping.

A standard approach to realize $U_{BP}$ in a quantum computing environment is adiabatic evolution, which slowly drives the system between the configurations encoded in $H_{ab}$. In this picture, each braiding step corresponds to an adiabatic interpolation between two Hamiltonians over a time interval $\tau$. The evolution from time $0$ to $\tau$ for a single step can be written as
\begin{equation}
    |\Psi(\tau)\rangle_\pm = e^{-i[(1-t)H_{12} + tH_{13}]}|\Psi(0)\rangle_\pm,
    \label{eq:AdiabaticEvolution}
\end{equation}
where $|\Psi(\tau)\rangle_\pm$ is the evolved state at time $\tau$, and $H_{12}$ and $H_{13}$ are the initial and final Hamiltonians for that step. The compact quantum circuit implementing the full braiding process via adiabatic evolution is shown in Fig.~\ref{fig:braid-adiabatic}; each operator in the circuit represents adiabatic evolution for a single step, subdivided into $\tau/\Delta t$ intervals. The resulting time-evolution operators are further trotterized to map them onto quantum gates.

For the experiments presented here, we employed the first-order Lie--Trotter decomposition~\cite{Berry2007} with a single repetition per trotterization step, further subdividing each interval into 10 time steps. This configuration was chosen to keep the total gate count in an experimentally reasonable range for the single-site trijunction benchmark, while maintaining an acceptable fidelity of the simulated braiding dynamics and balancing circuit depth with computational feasibility. In the absence of noise and Trotterization error, applying the inverse of the final state (denoted by $|\Psi\rangle_\mp^\dagger$ in Fig.~\ref{fig:braid-circuits}) would yield a unit probability of measuring $|000\rangle$. However, to keep the circuit depth minimal, we instead require that, even under noiseless simulation, a clearly dominant measurement probability for $|000\rangle$ suffices to confirm the process fidelity.

\subsection{Braiding Operator Based Emulation}
\label{Sec:Braiding}

While the adiabatic evolution operator in Section~\ref{Sec:Adiabatic} ensures smooth transitions between configurations, its reliance on trotterization renders it impractical for larger systems on NISQ hardware. To mitigate this, we introduce braiding operators that dramatically reduce circuit depth for braiding in a trijunction, where each arm of the trijunction contains $n$ sites. In this approach, the desired braid is implemented directly in the Majorana operator algebra, and only afterwards mapped to qubit operators via the chosen transformation.

A single braiding operator is constructed from multiple exchange operators, $\mathcal{O}_{kl}^{mp}$, which exchange the Majorana mode between site $k$ with orientation $m$ and site $l$ with orientation $p$~\cite{Ivano2001,Marek_2021}:
\begin{equation}
    \mathcal{O}_{kl}^{mp} = \frac{1}{\sqrt{2}}\left(1 + \gamma_k^m \gamma_l^p\right),
    \label{eq:BraidingOperator}
\end{equation}
satisfying the exchange relations
\begin{equation}
    \begin{split}
        \mathcal{O}_{kl}^{mp} \gamma_k^m \mathcal{O}_{kl}^{mp^\dagger} &= -\gamma_l^p,\\
        \mathcal{O}_{kl}^{mp} \gamma_l^p \mathcal{O}_{kl}^{mp^\dagger} &= \gamma_k^m,
    \end{split}
    \label{eq:BOProperties}
\end{equation}
so that they implement a non-trivial exchange of the corresponding Majorana modes.

The braiding operators for the trijunction system are obtained by sequencing these exchange operators. Each of the six braiding step from braiding protocol in figure \ref{fig:BraidingProtocol} can be further broken down into three exchange operator sub-steps.

\paragraph{Phase Change in Donor Arm}
The first sub-step begins with the topological phase in one of the donor arms, say $c$, changing to the trivial phase up to the nearest site next to the junction by exchanging Majoranas. We assume a clockwise direction for the motion of the trivial phase along the arms. The operator $U^{T}_{c}$ performs this sub-step for a single braiding step and can be defined as
\begin{equation}
    \begin{split}
        U^{T}_{c} &= \prod_{j=n-2}^{0} \mathcal{O}_{c(n-2-j)c(n-1-j)}^{yy} \\
        &= \prod_{j=n-2}^{0} \frac{1}{\sqrt{2}}\left(1 + \gamma_{c(n-2-j)}^y \gamma_{c(n-1-j)}^y\right).
    \end{split}
    \label{eq:UT}
\end{equation}

\paragraph{Junction Coupling Control}
The next sub-step exchanges Majorana modes at the junction between the nearest sites in the newly trivial arm $c$ and the existing trivial arm $a$. This operator transforms arm $c$ fully into the trivial phase, and the nearest site to the junction in arm $a$ now hosts the non-local Majorana. This operator $U^{J}_{ac}$ acts only on the sites nearest to the junction and is given by
\begin{equation}
    \begin{split}
        U^{J}_{ac} &=  \mathcal{O}_{c0a0}^{yy} \mathcal{O}_{c0a0}^{xx} \\
        &= \frac{1}{2}\left(1 + \gamma_{c0}^y \gamma_{a0}^y\right)\left(1 + \gamma_{c0}^x \gamma_{a0}^x\right).
    \end{split}
    \label{eq:UJ}
\end{equation}

\paragraph{Phase Change in Host Arm}
Finally, the last sub-step converts the phase of arm $a$ from trivial to topological. At this point, the non-local Majorana is hosted at the end of the now topological arm. This is performed via the operator $U^{R}_{a}$:
\begin{equation}
    \begin{split}
        U^{R}_{a} &= \prod_{j=n-2}^{0} \mathcal{O}_{a(j+1) a(j)}^{yy} \\
        &= \prod_{j=n-2}^{0} \frac{1}{\sqrt{2}}\left(1 + \gamma_{a(j+1)}^y \gamma_{a(j)}^y\right).
    \end{split}
    \label{eq:UR}
\end{equation}

\begin{figure}[t!]
  \centering
  \includegraphics[width=\linewidth]{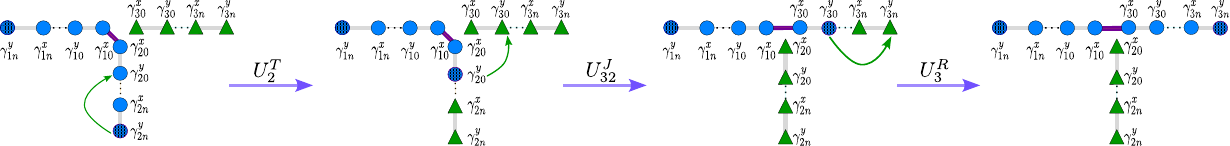}
  \caption{Sub-steps in a single braiding step of the braiding protocol. Green arrows indicate the motion of the Majorana zero mode (black square shading inside a blue circle), while violet arrows denote the order in which the braiding operators are applied. Green triangles represent Majorana bound states in the normal-phase wire, and blue circles represent Majorana bound states in the topological-phase wire.}
  \label{fig:SubStep}
\end{figure}
Six braiding operators are required to complete the full braiding protocol, and each braiding operator can be decomposed into the three sub-operators defined in~\eqref{eq:UT},~\eqref{eq:UJ}, and~\eqref{eq:UR}. Their respective effects are illustrated in Fig.~\ref{fig:SubStep}. For a trijunction with an arbitrary number of sites $n$ per arm, all three sub-operators contribute to each braiding step. In the minimal case of a single site per arm ($n=1$), the products in $U^T_{c}$ and $U^R_{a}$ are empty and these operators reduce to the identity, so each braiding step is implemented solely by the junction operator $U^J_{ac}$ acting on the three arm qubits, as reflected in the compact circuit shown in Fig.~\ref{fig:braid-operator}.

Depending on the mapping, the braiding operators can be transformed into Pauli-based operators using the appropriate transformation from Eqs.~\eqref{eq:JWTransform} or~\eqref{eq:CouplerTransform}. Standard decomposition methods then allow the Pauli-based operators to be compiled into quantum gates. For subsequent steps of the braiding protocol, the same operators apply, with only the arm indices changing as appropriate. The effect of each braiding operator on a given Hamiltonian is shown in Appendix~\ref{App:MQ_Braiding_Operators}.

\section{Results}
\label{Sec:Result}

This section presents numerical results assessing both the resource requirements and the physical correctness of the proposed braiding operator based emulation. First, we compare quantum circuit resources as a function of trijunction size and emulation method. We then verify that the resulting circuits implement the expected non-Abelian braiding transformation.

\subsection{Circuit Depth, Fidelity, and Two-Qubit Gate Counts}
\label{Sec:Result-resources}
Based on the mappings and simulation methods described above, we compare the circuit complexity of the adiabatic evolution and braiding operator based emulation approaches for increasing numbers of sites in each arm of the trijunction. The braiding operator are mapped to Pauli operators via the modified JW transformation, and then decomposed into native gate sequences as shown in fig.~\ref{fig:braid-operator}.

\begin{figure}[t!]
  \centering
  \includegraphics[width=0.99\linewidth]{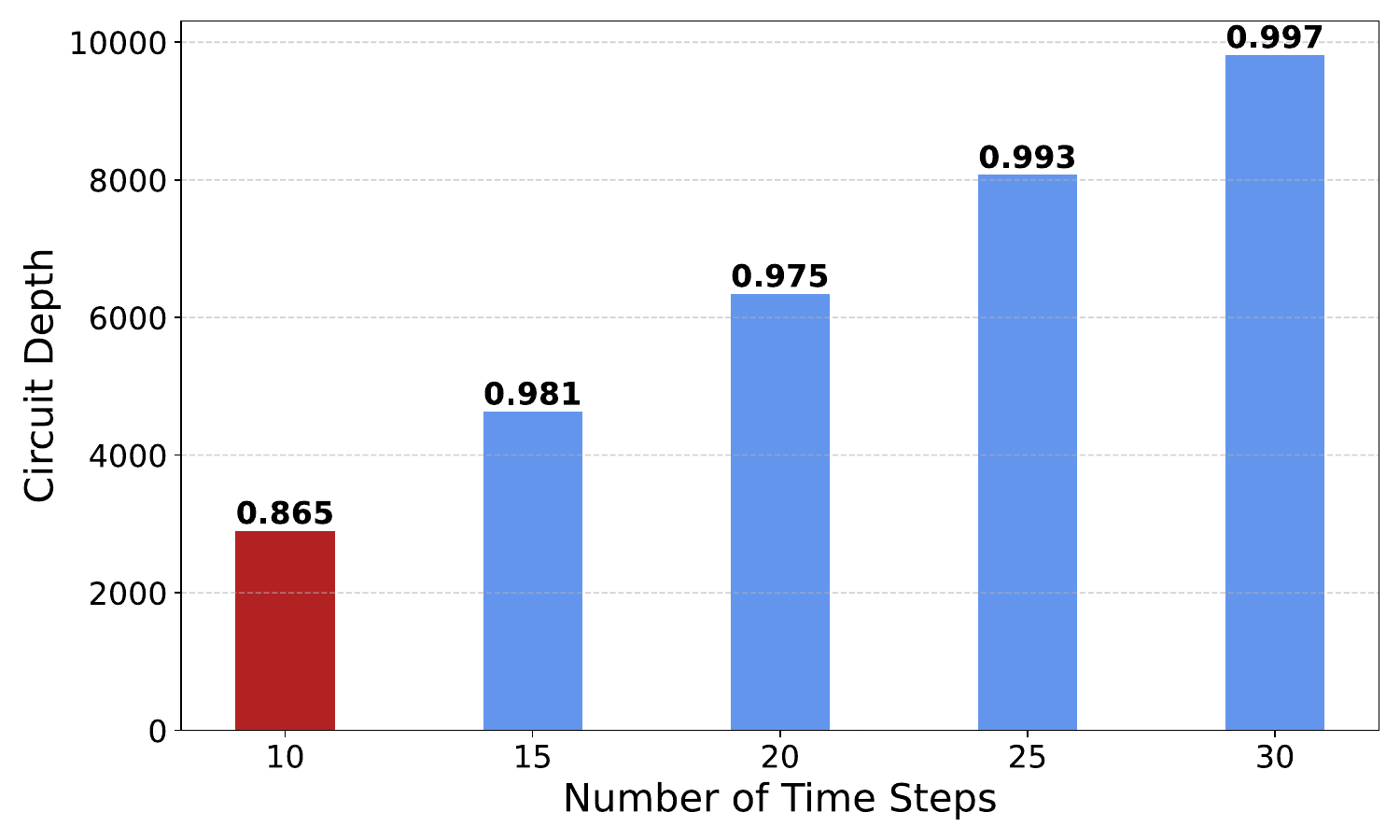}
  \caption{Trade-off between state fidelity and circuit depth for adiabatic simulation of a single-site trijunction. Each bar corresponds to the respective choice of Trotter steps in x-axis. The red bar indicates the minimal setting used in the resource comparison of fig.~\ref{fig:ECRComp}.}
  \label{fig:TrijunctionFidelityDepth}
\end{figure}

\begin{figure}[t!]
  \centering
  \includegraphics[width=0.99\linewidth]{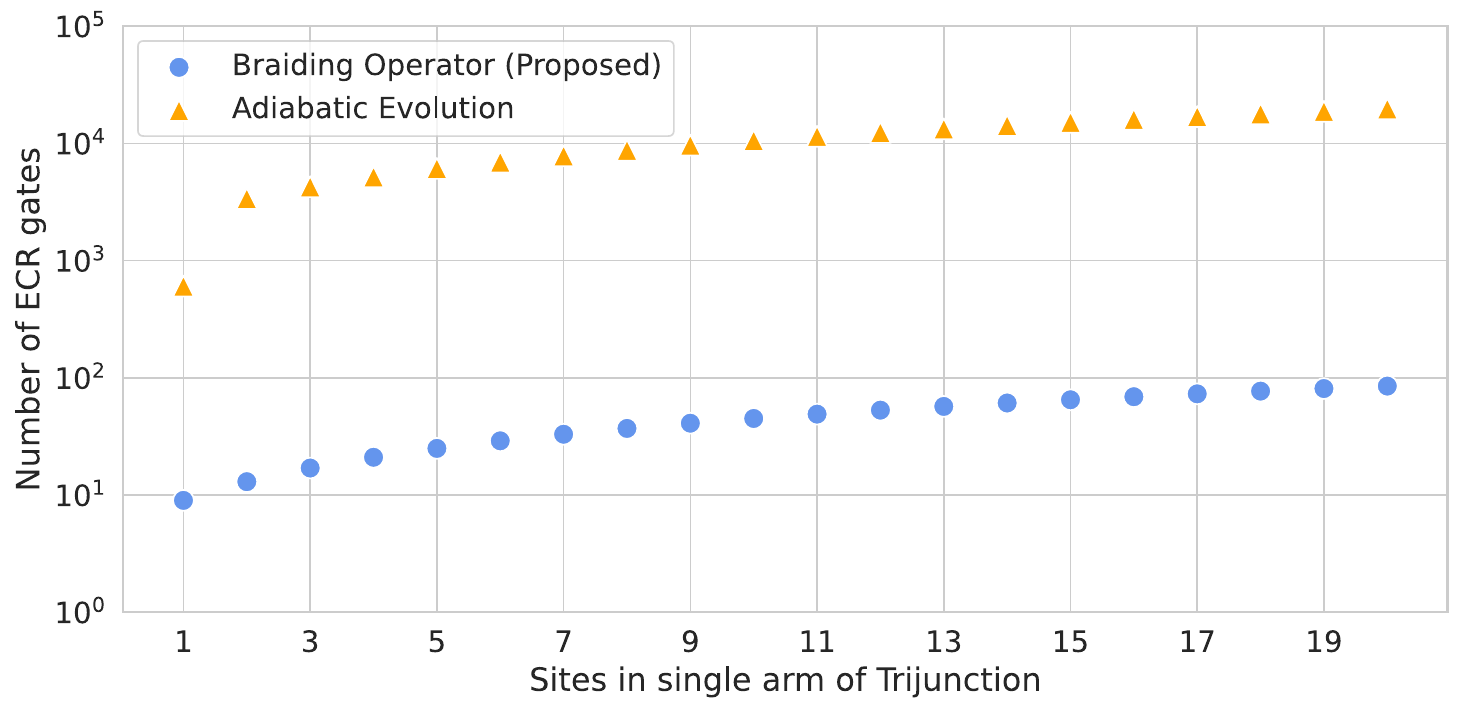}
  \caption{Total number of ECR gates for trijunctions with varying numbers of sites per arm (assuming all arms have equal sites) for comparison between methods of evolution: adiabatic evolution (orange triangles) and the braiding operator for coupler-based mapping (blue circles).}
  \label{fig:ECRComp}
\end{figure}

Figure~\ref{fig:TrijunctionFidelityDepth} shows the trade-off between state fidelity and circuit depth for adiabatic evolution of a single site trijunction using Lie--Trotter Trotterization. Increasing the number of Trotter steps improves the overlap with the ideal braiding evolution, but at the cost of a rapidly growing circuit depth and two-qubit gate count. For the resource comparison in fig.\ref{fig:ECRComp}, we therefore select the minimal Trotter setting, which keeps the ECR count as low as possible while still yielding acceptable fidelity for at least the single site benchmark. We deliberately use this same minimal setting for all trijunction sizes, rather than tuning the Trotter number to achieve a fixed fidelity (e.g., $>80\%$) for each size, in order to keep the total two-qubit gate count in a regime compatible with near-term experiments; enforcing a uniform high-fidelity threshold across all trijunction lengths would lead to a dramatic increase in circuit depth and ECR usage.

Figure~\ref{fig:ECRComp} then compares the resulting ECR counts for this adiabatic baseline against the braiding operator based emulation as the arm length is increased. The braiding-operator protocol exhibits much more favorable scaling in the number of sites per arm: for $n \ge 2$ it requires significantly fewer two-qubit gates than the adiabatic approach at the chosen minimal Trotter setting. This reduction is particularly important for NISQ devices, where ECR gates dominate both runtime and error accumulation.

\begin{figure}[t!]
  \centering
  \includegraphics[width=0.99\linewidth]{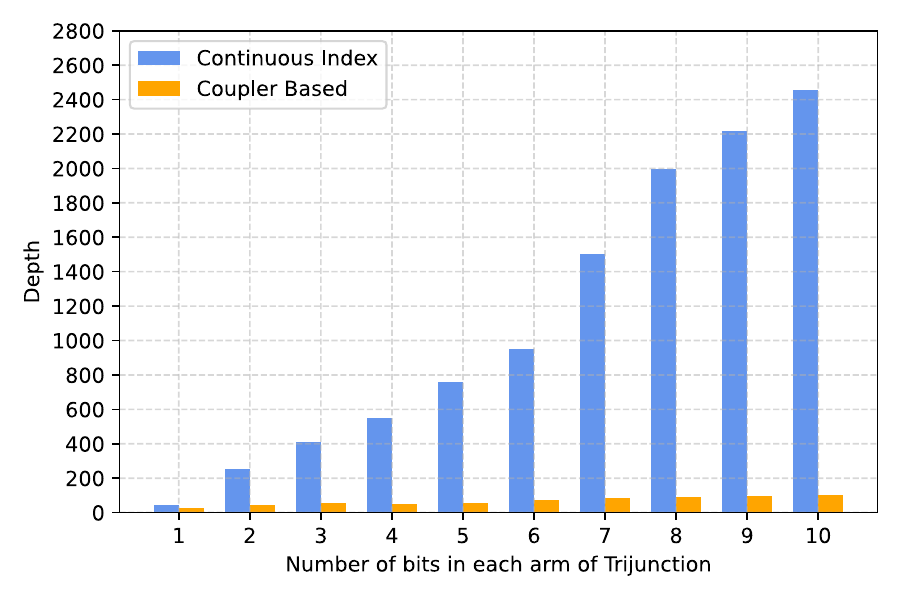}
  \caption{Circuit depth comparison for braiding-operator-based emulation using the coupler-based and continuous-index mappings. The coupler-based mapping yields substantially shallower circuits than the continuous-index mapping and is particularly well suited to devices with IBM’s heavy-hex qubit topology~\cite{Wootton2024}.}
  \label{fig:MappingComparision}
\end{figure}
Finally, Fig.~\ref{fig:MappingComparision} compares the overall circuit depth for the standard continuous index mapping and the coupler based mapping for trijunctions with varying numbers of sites per arm. The coupler based mapping consistently produces shallower circuits, reinforcing its advantage as the preferred mapping for implementing braiding operators on current superconducting hardware.

\subsection{Ground-State Braiding Phase}
\label{Sec:GS-phase}
In addition to reducing circuit depth, a valid braiding protocol must realize the characteristic non-Abelian transformation within the ground-state manifold. To verify this, we numerically examine the action of our braiding operators on the two-dimensional ground-state subspace of the trijunction Hamiltonian. We focus on the configuration in which arms $(1,2)$ are in the topological phase and arm $3$ is trivial, with Hamiltonian $H_{12}$. Let $\{\ket{g_1}, \ket{g_2}\}$ denote the two lowest-energy eigenstates of $H_{12}$, and define the isometry
\begin{equation}
    G = \bigl[\,\ket{g_1},\,\ket{g_2}\,\bigr],
\end{equation}
whose columns are the ground-state eigenvectors in the full Hilbert space. The effective braid operator acting on the ground-state manifold is then obtained by projecting the full unitary onto this subspace:
\begin{equation}
    U_{\mathrm{GS}} = G^\dagger U_{\mathrm{braid}} G,
    \label{eq:UGS-def}
\end{equation}
where $U_{\mathrm{braid}}$ is the unitary corresponding to a single braiding operation in our six-step protocol. Concretely, we take $U_{\mathrm{braid}}$ to be the evolution from $t = 0$ to $t = 3\tau$, where $\tau$ is the duration of one step in the protocol. At $t = 3\tau$, the trijunction returns to the same pattern of topological and trivial arms as at $t = 0$ (arms $1$ and $2$ topological, arm $3$ trivial), but the two nonlocal Majorana zero modes have swapped arms, corresponding to a single braiding operation in the trijunction geometry~\cite{Alicea2011,Clarke2011,Aasen2016}.

We evaluate $U_{\mathrm{GS}}$ for the minimal trijunction with a single site per arm and for a longer trijunction with three sites per arm. In both cases, $U_{\mathrm{GS}}$ is unitary up to numerical precision, with eigenvalues lying on the unit circle. After factoring out the global phase, the remaining eigenvalues differ by a relative phase $\Delta\phi \approx \pi/2$ for both the single-site and three-site trijunction. Equivalently, in the basis where $H_{12}$ defines the encoded fermionic mode $f = (\gamma_A + i\gamma_B)/2$, the two ground states acquire phases $e^{\mp i\pi/4}$ under $U_{\mathrm{braid}}$, in agreement with the standard Majorana braid operator $U_{AB} = \exp\!\bigl(\tfrac{\pi}{4}\gamma_A\gamma_B\bigr)$ acting on the encoded qubit~\cite{Ivano2001,Alicea2011,Aasen2016}. For the single-site trijunction, the explicit forms of $\ket{g_1}$, $\ket{g_2}$, and $U_{\mathrm{GS}}$ are given in Appendix~\ref{App:GS-phase}.

As a consistency check, we also consider applying the protocol twice in succession, corresponding to the full six-step evolution from $t = 0$ to $t = 6\tau$ and the unitary $U_{\mathrm{braid}}^2$. The projected operator $U_{\mathrm{GS}}^{(2)} = G^\dagger U_{\mathrm{braid}}^2 G$ then exhibits a relative phase $\Delta\phi \approx \pi$, consistent with the expected phase accumulation under two successive exchanges (a ``double braid'' or logical $Z$-type operation)~\cite{Ivano2001,Beenakker2020}. These results confirm that our braiding-operator-based protocol not only reduces the resource overhead, but also faithfully reproduces the non-Abelian braiding statistics of Majorana zero modes in the ground-state manifold of the trijunction.

\section{Conclusion}
\label{Sec:Conclusion}
We have introduced a scalable and resource-efficient method for emulating Majorana braiding in superconducting trijunction systems. By constructing braiding operators for each step of the braiding protocol, this operator-based approach enables emulation of braiding with a lower circuit depth and without requiring approximations to adiabatic evolution, thereby enhancing scalability. Our framework generalizes to trijunctions of arbitrary size linearly, offering a promising pathway for simulating topological quantum operations on near-term quantum devices.

\section{Continuous Index Mapping}
\label{App:JW-appendix}

The simplest mapping assigns the Kitaev chain sites linearly to qubits, with subsequent arms following sequentially. The general Majorana operators at site $j$ are expressed in terms of fermionic annihilation $(a_j)$ and creation operators $(a^\dagger_j)$ via:
\begin{equation}
    \begin{split}
        \gamma^x_j &= a_j + a^\dagger_j,\\
        \gamma^y_j &= -i(a_j - a^\dagger_j),
    \end{split}
    \label{eq:MajoranatoFermi}
\end{equation}
which are further transformed into Pauli operators using the Jordan--Wigner (JW) transformation~\cite{Jordan1928}:
\begin{equation}
    \begin{split}
        \gamma^x_j &= \sigma_{j-1}^x \prod_{i=0}^{j-2} \sigma_i^z, \\
        \gamma^y_j &= \sigma_{j-1}^y \prod_{i=0}^{j-2} \sigma_i^z.
    \end{split}
    \label{eq:JWTransform}
\end{equation}
Applying~\eqref{eq:JWTransform} and the Pauli commutation relations, we rewrite the system Hamiltonian in terms of Pauli operators:
\begin{equation}
    \begin{split}
        \mathcal{H}_{ab} &= -\Delta \sum_{k \in \{a, b\}} \sum^{n-2}_{j=0} \sigma_{n(k-1)+j}^x\sigma_{n(k1)+j+1}^x \\
        &\quad - \alpha \sum^{n-1}_{l=0} \sigma_{n(c-1)+l}^z \\
        &\quad + \epsilon_{abc} t_{ab} \sigma_{n(a-1)}^y\sigma_{n(b-1)}^x\prod^{n(b-1)-1}_{k = n(a-1)+1} \sigma_k^z.
    \end{split}
\end{equation}

\section*{Data Availability}
All data and code used in this study are available at \url{https://github.com/Rahps97/MZMBraidingEmulation.git}.

\section*{Acknowledgment}
R. S. gratefully acknowledges IBM Quantum and the Center for Quantum Information Physics at New York University for their support during the course of this research.

\appendix
\section{Multi-Qubit Braiding Operators}
\label{App:MQ_Braiding_Operators}
As a representative model to prove the use of braiding operators, we consider a trijunction of size \(3\), i.e., \(3\) sites in each arm. The trijunction in Fig.~\ref{fig:TriJunction} is the pictorial model for our proof and Fig.~\ref{fig:SubStep} represents the sub-steps. For times \(0\) to \(6\tau\), the system changes its configuration as depicted in Fig.~\ref{fig:BraidingProtocol}. The system is initialized in \(H_{12}\), derived from~\eqref{eq:MZMEquation} for a three-site trijunction (with \(\alpha = \Delta = t_{12} = 1\)),
\begin{equation}
    H_{12} 
    =  \sum_{k \in \{1, 2\}} \sum_{j=0}^{1} i\,\gamma_{kj}^y\gamma_{k,j+1}^x
      + \sum_{l=0}^{2} i\,\gamma_{3l}^x\gamma_{3l}^y
      + i\,\gamma_{10}^x\gamma_{20}^x .
    \label{eq:InitialThreeBits}
\end{equation}

The braiding operator for the first step in the protocol, \(T_1\), is obtained from the three sub-steps in Eqs.~\eqref{eq:UT}–\eqref{eq:UR} as
\begin{equation}
    T_{1} = U^R_3 U^J_{32} U^T_2.
    \label{eq:T1}
\end{equation}
To analyze its action on the Hamiltonian, it is convenient to introduce intermediate Hamiltonians. We first denote
\begin{equation}
    H^{(0)} \equiv H_{12},
\end{equation}
so that the full evolution can be written compactly as
\begin{equation}
    T_1 H_{12} T_1^\dagger
    = U^R_3 U^J_{32} U^T_2\,
      H^{(0)}\,
      U^{T\dagger}_2 U^{J\dagger}_{32} U^{R\dagger}_3.
    \label{eq:T1-conj-compact}
\end{equation}

Using~\eqref{eq:UT}, we can express the translation step as \(U^T_2 = \mathcal{O}^{yy}_{2021}\mathcal{O}^{yy}_{2122}\). Conjugating \(H^{(0)}\) by this unitary gives
\begin{equation}
  \begin{aligned}
    H^{(1)} &\equiv U^T_2 H^{(0)} U^{T\dagger}_2,\\
    H^{(1)} &= i \sum_{j=0}^{1} \gamma_{1j}^y \gamma_{1,j+1}^x
            + i \sum_{j=1}^{2} \gamma_{2j}^y \gamma_{2j}^x
            + i \sum_{j=0}^{2} \gamma_{3j}^x \gamma_{3j}^y\\
            &+ i\,\gamma_{10}^x \gamma_{20}^x .
  \end{aligned}
  \label{eq:H1}
\end{equation}
Next, the junction step \(U^J_{32}\) can be written in terms of local operators as \(U^J_{32} = \mathcal{O}^{yy}_{2030}\mathcal{O}^{xx}_{2030}\).
Conjugating \(H^{(1)}\) yields
\begin{equation}
  \begin{aligned}
    H^{(2)} &\equiv U^J_{32} H^{(1)} U^{J\dagger}_{32},\\
    H^{(2)} &= i \sum_{j=0}^{1} \gamma_{1j}^y \gamma_{1,j+1}^x
            + i \sum_{j=0}^{2} \gamma_{2j}^y \gamma_{2j}^x
            + i \sum_{j=1}^{2} \gamma_{3j}^x \gamma_{3j}^y\\
           & + i\,\gamma_{10}^x \gamma_{30}^x .
  \end{aligned}
  \label{eq:H2}
\end{equation}
Finally, the rotation step \(U^R_3\) can be written as \(U^R_3 = \mathcal{O}^{yy}_{3231}\mathcal{O}^{yy}_{3130}\). Conjugating \(H^{(2)}\) by \(U^R_3\) gives
\begin{equation}
  \begin{aligned}
    H_{13} &\equiv U^R_3 H^{(2)} U^{R\dagger}_3,\\
    H_{13} &= i\hspace{-0.25cm}\sum_{k \in \{1,3\}} \sum_{j=0}^{1}
              \gamma_{kj}^y \gamma_{k,j+1}^x
            + i\sum_{l=0}^{2}\gamma_{2l}^x \gamma_{2l}^y\\
            &- i\,\gamma_{10}^x \gamma_{30}^x .
  \end{aligned}
  \label{eq:H13}
\end{equation}
Combining~\eqref{eq:T1-conj-compact}–\eqref{eq:H13}, we obtain
\begin{equation}
    T_1 H_{12} T_1^\dagger = H_{13},
\end{equation}
which shows explicitly how the first braiding step maps the initial trijunction Hamiltonian \(H_{12}\) to \(H_{13}\).

The same procedure can be applied to each step of the braiding protocol. Evolving the system through all steps up to time \(3\tau\) returns the Hamiltonian to its original form at time \(0\). However, the braiding protocol only completes at \(6\tau\), when the two distinguishable Majorana modes return to their initial sites, having undergone a non-trivial exchange.

\section{Ground-State Braid Operator}
\label{App:GS-phase}
For the minimal trijunction with a single site per arm, the Hilbert space is four qubits (three arm qubits plus the coupler), and the braiding unitary $U_{\mathrm{braid}}$ obtained from the six-step protocol is a $16\times 16$ matrix in the computational basis $\{\ket{q_3 q_2 q_1 q_0}\}_{q_i\in\{0,1\}}$. In this appendix we show explicitly how the braid acts on the two-dimensional ground-state subspace.
For the configuration with arms $(1,2)$ in the topological phase and arm $3$ trivial, the two lowest-energy eigenstates of $H_{12}$ are
\begin{equation}
    \begin{split}
        \mid g_1 \rangle &= \frac{1}{\sqrt{2}}\bigl(\ket{0000} + \ket{0110}\bigr),\\
        \mid g_2 \rangle &= \frac{1}{\sqrt{2}}\bigl(\ket{0010} + \ket{0100}\bigr),
    \end{split}
    \label{eq:gs-basis-1site}
\end{equation}
which form an orthonormal basis of the ground-state manifold. It is convenient to restrict $U_{\mathrm{braid}}$ to the four-dimensional subspace spanned by $\{\ket{0000},\ket{0010},\ket{0100},\ket{0110}\}$. In this ordered basis, the unitary corresponding to a single braid (the evolution from $t=0$ to $t=3\tau$) has the block form
\begin{equation}
    U_{\mathrm{sub}}
    =
    \begin{pmatrix}
        1 & 0 & 0 & 0 \\
        0 & 0 & i & 0 \\
        0 & i & 0 & 0 \\
        0 & 0 & 0 & 1
    \end{pmatrix},
    \label{eq:Usub-1site}
\end{equation}
where all other basis states decouple and transform trivially.
Acting with $U_{\mathrm{sub}}$ on the ground states in Eq.~\eqref{eq:gs-basis-1site} gives
\begin{equation}
    \begin{split}
        U_{\mathrm{sub}}\ket{g_1}
        &= \frac{1}{\sqrt{2}}\bigl(U_{\mathrm{sub}}\ket{0000}
                                  + U_{\mathrm{sub}}\ket{0111}\bigr)
         = \ket{g_1}, \\
        U_{\mathrm{sub}}\ket{g_2}
        &= \frac{1}{\sqrt{2}}\bigl(U_{\mathrm{sub}}\ket{0010}
                                  + U_{\mathrm{sub}}\ket{0100}\bigr)
         = i\,\ket{g_2},
    \end{split}
    \label{eq:gs-action-1site}
\end{equation}
since Eq.~\eqref{eq:Usub-1site} implies
$U_{\mathrm{sub}}\ket{0000} = \ket{0000}$,
$U_{\mathrm{sub}}\ket{0010} = i\ket{0100}$,
$U_{\mathrm{sub}}\ket{0100} = i\ket{0010}$, and
$U_{\mathrm{sub}}\ket{0110} = \ket{0110}$.
In the $\{\ket{g_1},\ket{g_2}\}$ basis, the projected braid operator
\begin{equation}
    U_{\mathrm{GS}} = G^\dagger U_{\mathrm{braid}} G,\qquad
    G = \bigl[\,\ket{g_1},\,\ket{g_2}\,\bigr],
\end{equation}
is therefore diagonal:
\begin{equation}
    U_{\mathrm{GS}}
    =
    \begin{pmatrix}
        1 & 0 \\
        0 & i
    \end{pmatrix}.
    \label{eq:UGS-1site}
\end{equation}
The eigenvalues $\{1,i\}$ differ by a relative phase
$\Delta\phi = \pi/2$, in agreement with the standard Majorana braid
operator $U_{AB} = \exp\!\bigl(\tfrac{\pi}{4}\gamma_A\gamma_B\bigr)$
acting on the encoded fermionic mode~\cite{Ivano2001,Alicea2011,Aasen2016}.

Repeating the protocol twice corresponds to the unitary
$U_{\mathrm{braid}}^2$, whose restriction to the same four-dimensional
subspace is simply
\begin{equation}
    U_{\mathrm{sub}}^2
    =
    \begin{pmatrix}
        1 & 0 & 0 & 0 \\
        0 & -1 & 0 & 0 \\
        0 & 0 & -1 & 0 \\
        0 & 0 & 0 & 1
    \end{pmatrix},
\end{equation}
so that in the ground-state basis we obtain
\begin{equation}
    U_{\mathrm{GS}}^{(2)}
    = G^\dagger U_{\mathrm{braid}}^2 G
    =
    \begin{pmatrix}
        1 & 0 \\
        0 & -1
    \end{pmatrix},
    \label{eq:UGS2-1site}
\end{equation}
with a relative phase $\Delta\phi = \pi$ between the two ground states.
This is consistent with the expected phase accumulation for two
successive exchanges (a double braid) in the Majorana encoding.

\bibliographystyle{IEEEtran}
\bibliography{citation}

\end{document}